\documentclass[11pt,letterpaper]{article}
\usepackage{mathrsfs}[10pt]
\usepackage{url}
\usepackage{amsmath}
\usepackage{url}
\usepackage[dvips]{graphicx}
\usepackage{float}
\usepackage[numbers,sort&compress]{natbib}
\usepackage{algorithm}
\usepackage{algorithmic}
\usepackage{multirow}
\usepackage{fancyhdr}
\usepackage{threeparttable}
\usepackage{subfigure}

\begin{document}

\title{Protein Inference and Protein Quantification: Two Sides of the Same Coin}
\author{\\Ting Huang, Peijun Zhu, Zengyou He\footnote{Corresponding author}}
\date{School of Software, Dalian University of Technology, Dalian, China\\
\emph{thuang0703@gmail.com, zhupeijun@outlook.com, zyhe@dlut.edu.cn}
}
\maketitle
\date{}

\thispagestyle{empty}

\begin{abstract}
\noindent\emph{Motivation}: In mass spectrometry-based shotgun proteomics, protein quantification and protein identification are two major computational problems. To quantify the protein abundance, a list of proteins must be firstly inferred from the sample. Then the relative or absolute protein abundance is estimated with quantification methods, such as spectral counting. Until now, researchers have been dealing with these two processes separately. In fact, they are two sides of same coin in the sense that truly present proteins are those proteins with non-zero abundances. Then, one interesting question is if we regard the protein inference problem as a special protein quantification problem, is it possible to achieve better protein inference performance? \\

\noindent\emph{Contribution}: In this paper, we investigate the feasibility of using protein quantification methods to solve the protein inference problem. Protein inference is to determine whether each candidate protein is present in the sample or not.  Protein quantification is to calculate the abundance of each protein. Naturally, the absent proteins should have zero abundances. Thus, we argue that the protein inference problem can be viewed as a special case of protein quantification problem: present proteins are those proteins with non-zero abundances. Based on this idea, our paper tries to use three very simple protein quantification methods to solve the protein inference problem effectively.\\

\noindent\emph{Results}: The experimental results on six datasets show that these three methods are competitive with previous protein inference algorithms. This demonstrates that it is plausible to take the protein inference problem as a special case of protein quantification, which opens the door of devising more effective protein inference algorithms from a quantification perspective. \\

\noindent\emph{Availability}: The source code of our methods is available at:
\url{http://code.google.com/p/protein-inference/}\\

\end{abstract}
\newpage
\setcounter{page}{1}

\section{Introduction}

\begin{figure}[!tpb]
\centerline{\includegraphics[scale=0.6]{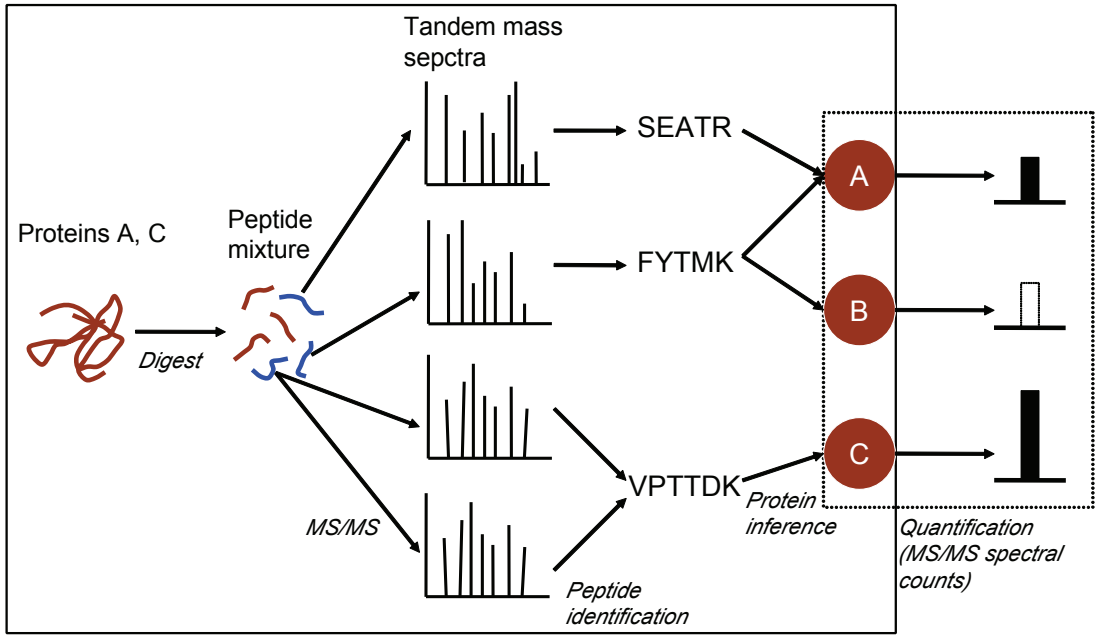}}
\caption{Protein identification and quantification using mass spectrometry in shotgun proteomics.  There are three major computational problems: peptide identification, protein inference and protein quantification.}\label{fig1}
\end{figure}

Mass spectrometry (MS)-based shotgun proteomics is currently the most widely used method
for the identification and quantification of proteins \cite{Analysis}. As shown in Fig.\ref{fig1}, it first digests the sample into a mixture of peptides by enzymes such as trypsin.  The resulting peptide mixtures are scanned by tandem mass spectrometry (MS/MS) to generate a set of MS/MS spectra. Then the peptide search engine reports a set of peptide-spectrum matches (PSMs) by searching the MS/MS spectra against a protein database. From these peptide identifications, we infer the existence of proteins with protein inference algorithms and calculate the relative or absolute abundances of proteins with protein quantification approaches \cite{Review,Nikolov}.

Until recently, people tackle the identification and quantification of proteins as two individual
and subsequent tasks: first select a subset of proteins that are truly present and then determine the
abundances of these proteins. For both problems, many elegant approaches \cite{DTASelect,Dbparser,Proteomicbing,Improvedze,Markey,Aseon,Ebpthomas,Probabilityfeng,Aweaherly,Ayong,Efficientoliver,Anmoore,Statisticalr,Improved,Aqunhua,Achangyu,Direct,Searle,Azeng,Improvingbingwen,Proteinpaul,Ramakrishnan1,Networkjing,Ramakrishnan2,Grobei,Douglas,Proteinsarah,William,Florens,Recomb2009SharedPeptide}
have been developed in the past decades. The readers can refer to two recent reviews \cite{Review,Nikolov} for details.

The starting point of this paper is the observation of some key underlying connections between
these two problems. In protein inference, the objective is to generate a binary presence indicator
value (1 or 0) for each candidate protein. In this regard, ``protein existence inference" is probably
more accurate in describing the original protein inference task. In protein quantification or ¡°protein
abundance inference¡±, the objective is to determine the abundances of a set of proteins. Clearly, if
one protein is not present, its abundance should be 0. Hence, we argue that the protein inference
problem can be investigated from the perspective of protein quantification: present proteins are
those proteins with non-zero abundances. In other words, we can adopt available protein quantification
methods directly to solve the protein inference problem. This new angle may enable a better
understanding of the protein inference problem and help in devising improved or hybrid methods
by combining elements from two areas that would otherwise be considered incompatible.

As a proof of concept, we investigate the feasibility of solving protein inference problem with
existing protein quantification methods in the context of label-free proteomics. In label-free quantitative proteomics studies, quantification methods based on peak ion intensities (from MS data) \cite{Muralidharan} and spectral counting (from MS/MS data) \cite{Linfeng,Fermin} have been widely used.

Spectral counting measures the abundance of each protein based on the number of MS/MS spectra that match its constituent peptides. Compared to peptide intensity values, spectral counting information is easier to obtain since we just need to count the number of the MS/MS spectra. In this paper, we use spectral counting as the quantification approach for solving the protein inference problem.

We first try two simple spectral counting methods in the literature. In both methods, the protein abundance is calculated as the sum of peptide abundance. Their difference lies in how to handle the shared peptide. If the abundance of one shared peptide is $b$ and it has $k$ parent proteins, then $b$ is used as its abundance in the first method while $b/k$ is used in the second method.  These two methods assume that all the candidate protein are present in the sample and they have non-zero abundances. However, this assumption contradicts the objective of protein inference: distinguish present proteins with non-zero abundances from absent proteins with zero abundances. Thus, we come up with another linear programming model to shrink some protein abundances to zero.

To our knowledge, our paper is the first attempt to use protein quantification methods for protein inference. Such an attempt connects two important computational problems that have long been investigated separately.  The experimental results show that we can obtain better performance in most datasets even when the most simple version of spectral counting is utilized.  Hence, the advance in protein quantification studies will promote the development of more effective protein inference algorithms.

In Section 2, we describe the details of three methods. Section 3 shows the experimental
results on six datasets. Section 4 concludes the paper.

\section{Methods}

\begin{figure}[!tpb]
\centerline{\includegraphics[scale=0.5]{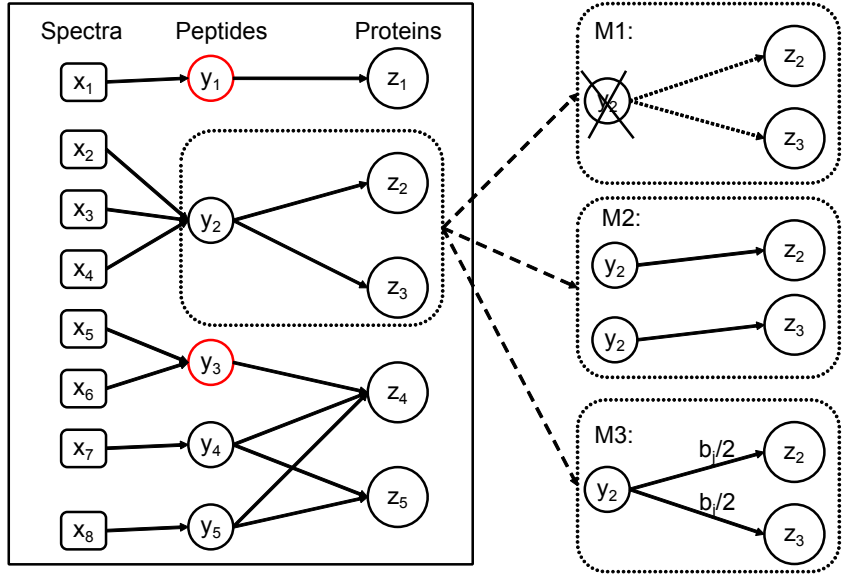}}
\caption{Three approaches used in the spectral counting for solving shared peptide problem. $y_1$ and $y_3$ are unique peptides while $y_2$, $y_4$ and $y_5$ are shared peptides. The abundance of peptide $y_j$ is represented by $b_j$. We use peptide $y_2$ as an example to explain how these three approaches work. }\label{fig2}
\end{figure}

As shown in the left side of Fig.\ref{fig2}, the input of the protein inference problem can be represented as a tripartite graph $G = (X\cup{Y\cup{Z}},E_1\cup{E_2})$ where X, Y and Z are the set of $l$ MS/MS experimental spectra, $m$ identified peptides and $n$ candidate proteins respectively. For all $x_i\in{X}$, $y_j\in{Y}$, there is an edge $(x_i,y_j)\in{E_1}$ if and only if spectrum $x_i$ matches the peptide $y_i$ in the peptide identification results. Similarly, $(y_j,z_k)\in{E_2}$ means that peptide $y_j$ is one part of the protein sequence $z_k$ . Each MS/MS spectrum corresponds to one and only one identified peptide whereas some peptides may have more than one matching spectrum, such as peptide $y_2$ and $y_3$ in Fig.\ref{fig2}. The relationship between the peptides and proteins is more complex: candidate proteins may have several identified peptides and peptides can be shared by multiple proteins. How to correctly distribute these shared peptides is one of the most challenging problem in protein inference.

We first formulate the protein inference problem as a special case of protein quantification problem. The objective of protein inference is to determine whether each candidate protein is present in the sample. The aim of protein quantification is to estimate the abundances of a set of proteins. Clearly, if one protein is not present, its abundance should be 0. In this paper, the protein inference problem is re-visited from the perspective of protein quantification through seeking those proteins with non-zero abundances.

To obtain the protein abundance,  we start with calculating the peptide abundance. The abundance $b_j$ of peptide $y_j$ is calculated as the sum of PSM  probabilities (or scores):
\begin{equation}
b_j=\sum_{(x_i,y_j)\in{E_1}}a_i,\label{eq:01}
\end{equation}
where $a_i$ is the probability that spectrum $x_i$ matches peptide $y_j$.  $a_i$ can be also viewed as the weight of edge $(x_i,y_j)\in{E_1}$, which can be obtained from peptide identification algorithms such as Mascot \cite{Probability} or post-processing tools such as PeptideProphet \cite{Aebersold3}. In traditional spectral counting methods, the peptide abundance is simply the number of MS/MS spectra identified for each peptide. Here, we generalize this spectral counting method to account for the quality of PSMs.  More precisely, the contribution of each spectrum to the peptide abundance is a quantitative value between 0 and 1 rather than a fixed value of one.  Such an extension is extremely important for protein inference since it may help us to distinguish proteins with the same number of PSMs.

To calculate the protein abundance, we need to distribute the abundance of each peptide to its parent proteins. The main difficulty is how to deal with the degenerate peptide that is shared by more than one protein since such a peptide can be generated by any subset of its parent proteins.

There are several approaches to solve the shared peptide problem in protein quantification \cite{Florens,Usaite,Zybailov}, as shown in the right side of Fig.\ref{fig2}. The first approach is to simply discard the shared peptides and only use the unique peptides to calculate the protein abundance. But this approach has one disadvantage: it causes the information lost, especially for proteins whose identified peptides are all shared peptides. In Fig.\ref{fig2}, if we delete the shared peptide $y_2$, then proteins $z_2$ and $z_3$ don't have any identified peptide and they would be considered as being absent in the sample. In fact, at least one of these two proteins must be present if we assume the existence of peptide $y_2$.  Alternatively, we can use both unique and shared peptides to estimate the protein abundance. In the second approach, the abundance of each shared peptide is utilized in the abundance calculation of all its parent proteins. In other words, each peptide is counted multiple times so that the abundances of some proteins may be over-estimated. We call this method ``multiple counting" in this paper. For example, peptide $y_2$ in Fig.\ref{fig2} is counted twice in the second approach, which means that we artificially increase the abundance of peptide $y_2$ from $b_2$ to $2*{b_2}$. The third approach divides the abundance of one shared peptide into different parts and then distributes each part to one of its parent proteins.  This approach ensures that each peptide is ``counted" only once. One typical representative in this category is the ``equal division" method, which partitions the peptide abundance into $k$ equal parts ($k$ is the number of proteins that share this peptide).

Since both multiple counting and equal division are the most popular and simple approaches for spectral counting based protein quantification, we first try these two methods for protein inference and see how they perform.  Both of these two methods assume that all the candidate proteins are present in the sample and should have non-zero abundances. However, this assumption doesn't hold in protein inference because some absent proteins should have zero abundances. Thus,  a new linear programming model is proposed as well to distribute peptide abundance automatically and set the abundances of some proteins to be zero.

\subsection{Multiple Counting}
In this method, shared peptides are used in the same way as the unique peptides and receive no special treatment. The protein abundance is simply the sum of peptide abundance from both shared and unique peptides corresponding to each protein:
\begin{equation}
c_k=\sum_{(y_j,z_k)\in{E_2}}b_j,\label{eq:02}
\end{equation}
where $c_k$ is the abundance of protein $z_k$. If peptide $y_j$ has $q_j$ parent proteins, then it is counted $q_j$ times and its actual abundance used in the calculation is $q_j*{b_j}$.

\subsection{Equal Division}
Different from the above method that counts shared peptides multiple times, this method counts each peptide only once. It equally distributes the abundance of each shared peptide to its parent proteins:
\begin{equation}
c_k=\sum_{(y_j,z_k)\in{E_2}}\frac{b_j}{q_j},\label{eq:03}
\end{equation}
where $q_j$ is the number of candidate proteins sharing peptide $y_j$. If peptide $y_j$ is a unique peptide, then $q_j=1$.

\subsection{Linear Programming Model}

For each identified peptide $y_j$, the peptide abundance can be computed as:
\begin{equation}
b_j=\sum_{\{k|(y_j,z_k)\in{E_2}\}}d_{jk},\label{eq:04}
\end{equation}
where $d_{jk}$ can be interpreted as the abundance that protein $z_k$ contributes to peptide $y_j$. The variable $d_{jk}$ can serve as the bridge between peptide
abundance and protein abundance. On one hand, we can use $d_{jk}$ to explain the known peptide abundance. On the other
hand, we can calculate the unknown protein abundance through $d_{jk}$. Therefore, the protein quantification based protein inference problem is equivalent to find an optimal matrix $D = (d_{jk})$.

According to the above analysis, we propose a linear programming (LP) model to solve the protein inference problem:
\begin{eqnarray}
\label{eq:05}\min_{D}{\sum_{k=1}^{n}{t_k}}\\
\label{eq:06}\forall{j,k}:d_{jk}\leq{t_k}\\
\label{eq:07}\forall{j}:b_{j}-\sum_{\{k|(y_j,z_k)\in{E_2}\}}{d_{jk}}=0\\
\label{eq:09}\forall{j,k}:d_{jk}\sim
\begin{cases}
={0}\quad{if\ \displaystyle{(y_j,z_k)\notin{E_2}}}\\
\geq0\quad{\ \displaystyle{else}}
\end{cases}.
\end{eqnarray}

Some further illustrations on the model are listed as follows:
\begin{itemize}
\item Constraint (\ref{eq:06}) is to find the maximum value in each column vector $d_k$  (the $k$th column of matrix $D$). Since we regard the proteins with non-zero abundances as being present in the sample, the abundances of absent proteins should be zero. Therefore, we minimize the sum of maximum peptide abundance from each protein in the objective function so as to shrink some protein abundances to 0.

\item The left-hand side of constraint (\ref{eq:07}) is the difference between the observed and predicted peptide abundance. $b_j$ is viewed as the observed value and the sum of $d_{jk}$ is the predicted value. The zero difference means that each peptide should be ``counted" only once in distributing the abundance of shared peptide.

\item  In constraint (\ref{eq:09}), we set $d_{jk}=0$ if $(y_j,z_k)\notin{E_2}$ and consider only the remaining elements of matrix $D$ as variables. This greatly improves the running efficiency of the LP model.

\item Dost et al. \cite{Recomb2009SharedPeptide} have presented a similar LP model in their $F_2$ formulation. It aims at inferring the protein abundance and peptide detectability simultaneously. The biggest difference between these two LP models is that our model sets some protein abundances to 0 while Dost's method doesn't.
\end{itemize}

After obtaining the matrix $D$, the protein abundance $c_k$ is calculated as:
\begin{equation}
c_k=\sum_{\{j|(y_j,z_k)\in{E_2}\}}{d_{jk}}.\label{eq:10}
\end{equation}

\subsection{Converting Scores into Probabilities}
After knowing the protein abundance, it is beneficial to convert the abundance into well-calibrated probability. The main reason is that the probability estimation allows us to select the appropriate threshold for reporting the present proteins. In fact, the problem of converting ranking scores into estimated probabilities has been widely investigated in different domains (e.g., \cite{JingGao}). In this paper, we use the method proposed in \cite{JingGao} to fulfill this task.

We first estimate the probability $p_k$ that protein $z_k$ is present in the sample given its abundance $c_k$:
\begin{equation}
Pr(z_k=1|c_k) = \frac{Pr(c_k|z_k=1)Pr(z_k=1)}{Pr(c_k|z_k=1)Pr(z_k=1)+Pr(c_k|z_k=0)Pr(z_k=0)}\\
=\frac{1}{1+\exp(-f_k)},\label{eq:11}
\end{equation}
where
\begin{equation}
f_k = \log\frac{Pr(c_k|z_k=1)Pr(z_k=1)}{Pr(c_k|z_k=0)Pr(z_k=0)}.\label{eq:12}
\end{equation}

Assuming $f_k$ has a Gaussian distribution with equal covariance matrices, Equation (\ref{eq:11}) becomes
\begin{equation}
p_k=\frac{1}{1+\exp(A{c_k}+B)}.\label{eq:13}
\end{equation}

Now, we need to learn the parameters, $A$ and $B$. Let $r_k$ be a binary variable whose value is 1 if protein $z_k$ is present in the sample and 0 otherwise. Then, $R =(r_1,r_2,\cdots,r_n)$ is the presence indicator vector of $n$ candidate proteins. If we assume that the existence of each protein is independent with other proteins, the probability of observing $R$ given $C$ is:
\begin{equation}
Pr(R|C)=\sum_{k=1}^{n}p_k^{r_k}{(1-p_k)^{1-r_k}},\label{eq:14}
\end{equation}
where $C=\{c_1,c_2,\cdots,c_n\}$. The optimal parameter values should maximize $Pr(R|C)$, i.e., minimize the following negative log likelihood function:
\begin{equation}
LL(R|C)=\sum_{k=1}^{n}{[(1-r_k)(-A{c_k}-B)+\log(1+\exp(A{c_k}+B))]}.\label{eq:15}
\end{equation}

Equation (\ref{eq:15}) is based on the assumption that we have already known the indicator vector $R$. However, we don't know such information in the protein inference process.  Thus, we consider $r_k$s as hidden variables and employ an EM algorithm \cite{JingGao} to simultaneously estimate $A$, $B$ and $R$.

The EM algorithm utilizes an iterative procedure to estimate the parameter values $\theta=\{A,B\}$. The procedure includes two steps: set $r_{k}^{s+1}=E(r_k^s|C,\theta^{s})$ (E-step) and compute $\theta^{s+1}=\arg\min_{\theta}LL(R^{s+1}|C)$ (M-step) where $s$ is the iteration index. During the E-step, the unknown vector $R$ is replaced by its expected value $R^{s+1}$ under the current estimated parameter values $\theta^{s}$. Since $\theta^{s}$ are fixed, $LL(R|C)$ is minimized by setting $r_k=0$ if $Ac_k+B>0$ or $r_k=1$ if $Ac_k+B\leq{0}$. During the M step, a new parameter estimation $\theta^{s+1}$ is computed by minimizing $LL(R|C)$ given the $R^{s+1}$ values  calculated by the first step. Since $R^s =[r^s_k]$ is fixed, minimizing $LL(R|C)$ with respect to $A$ and $B$ is a two-parameter optimization problem, which can be solved using the model-trust algorithm described in \cite{Platt}.

\section{Experimental Results}
To test the performance of our methods, we have compared them with ProteinProphet \cite{Aseon} and MSBayesPro \cite{Ayong}  on six datasets.

\subsection{Datasets}

We use six datasets that are publicly available and their URLs are given in Table~1.  Among these six datasets, 18 mixtures \citep{18mixtures}, Sigma49 and yeast \citep{Ramakrishnan1} have a corresponding protein reference set as the set of ground-truth proteins. An identified protein is labeled as a true identification if it is present in the protein reference set. Another three datasets, DME \citep{DMEBrunner}, HumanMD \citep{Ramakrishnan2} and HumanEKC \citep{Ramakrishnan1}, have no such sets. Thus, we use a target-decoy strategy for performance evaluation, in which the MS/MS spectra are searched against a mixed protein database containing all target protein sequences and an equal number of decoy sequences. Using this strategy, an identified protein is considered as a true identification if it comes from the target protein database.

\textbf{Mixture of 18 Purified Proteins (18 mixtures).}
The first dataset is a synthetic mixture of 18 highly purified proteins from \textit{ISB Standard Protein Mix Database} \cite{18mixtures}. The protein database consists of 1819 protein sequences including 18 standard proteins with contaminant entries appended to the database.

\textbf{Sigma49 Dataset.} Sigma49 is a synthetic mixture of 49 human proteins. The database used for peptide identification is composed of 15682 Swiss-Prot human protein sequences.

\textbf{Yeast Dataset.}
This dataset has been used in \cite{Ramakrishnan1}. The reference set is generated by an intersection of identified proteins from 4 MS-based proteomics datasets and 3 non-MS-based datasets. It contains 4265 proteins observed in either two or more MS datasets or any of non-MS datasets and is available at \url{http://www.marcottelab.org/MSdata/gold/yeast.html}. The database used in the experiment contains 6,714 protein sequences.

\textbf{D. melanogaster Dataset (DME).}
DME comes from the embryonal Kc 167 cell line of D. melanogaster \cite{DMEBrunner}. Its corresponding protein database is the release 5.2 from Flybase with 20,726 entries.

\textbf{HumanMD Dataset.}
This dataset has been used in \cite{Ramakrishnan2}. Its sample is a medulloblastoma Daoy cell line obtained from American Type Culture Collection (ATCC).  The protein database is Ensembl version 49.36k with 22,997 entries.

\textbf{HumanEKC Dataset.}
HumanEKC has been used in \cite{Ramakrishnan1}. It is generated from a human embryonic kidney T293 cell line of ATCC. Its database is the same as that of HumanMD dataset.
\begin{table*}[htbp]
\centering
\label{tab:decoydataset}
\caption{{\footnotesize\textbf{Dataset URL.}}}
\begin{tabular}{c|c}\hline
\footnotesize\textbf{Dataset}&\footnotesize\textbf{Raw data URL}\\\hline
\footnotesize{Mixture of 18 Purified Proteins} &
\footnotesize{\url{http://regis-web.systemsbiology.net/PublicDatasets/}}\\\hline
\footnotesize{Sigma49 Dataset}&
\footnotesize{\url{https://proteomecommons.org/dataset.jsp?i=71610}}\\\hline
\footnotesize{Yeast Dataset}&
\footnotesize{\url{http://www.marcottelab .org/users/MSdata/Data_02/}}\\\hline
\footnotesize{D. melanogaster Dataset} &
\footnotesize{\url{http://www.peptideatlas.org/repository/} (PAe001349)}\\\hline
\footnotesize{HumanMD Dataset}&
\footnotesize{\url{http://www.marcottelab.org/MSdata/Data_05/}}\\\hline
\footnotesize{HumanEKC Dataset}&
\footnotesize{\url{http://www.marcottelab.org/MSdata/Data_07/}}\\\hline
\end{tabular}
\end{table*}
\subsection{Peptide Identification}

We use X!Tandem (v2010.10.01.1) \citep{Unimod} as the peptide identification software. For 18 mixtures, Sigma49 and yeast datasets, all MS/MS spectra are only searched against the target protein databases. For DME, HumanMD and HumanEKC datasets, the spectra need to search against both target and decoy protein databases.

During the database search, we use default search parameters. Then, peptide-spectra matching probabilities are computed using PeptideProphet included in Trans-Proteomic Pipeline (TPP) v4.5.

\subsection{Protein Inference}

We compare our methods with  ProteinProphet and MSBayesPro. ProteinProphet is the most popular method for protein inference so far. MSBayesPro is one representative of recently proposed methods and its software package is publicly available. We run ProteinProphet with its default parameter setting. MSBayesPro uses peptide detectability information as one part of its input. For some peptides whose detectabilities cannot be predicted by the current software, we calculate them by ourselves: the detectability value = median(predicted detectability scores from the same parent protein)/3.

For the proteins that cannot be distinguished with respect to identified peptides, ProteinProphet and our LP model put all of them into the same group. Whenever we refer to the number of true positives (TPs) or false positives (FPs) identified at a threshold or use these values in a calculation, all proteins in the group are reported and the group probability is used as their protein probabilities.

\subsection{Results}

We evaluate the performance of different methods using a curve that plots the number of TPs as a function of $q$-value. An identified protein is labeled as a TP if it is present in the protein reference set or target protein sequence database, and as a FP otherwise. Given a certain probability threshold $t$, suppose there are $T_{t}$ TPs and $F_{t}$ FPs, the false discovery rate (FDR) is estimated as $FDR_{t} = F_{t}/(F_{t}+T_{t})$. The corresponding $q$-value is defined as the minimal FDR that a protein is reported: $q_t=\min_{t^{'}\leq{t}}FDR_{t^{'}}$.  The curve is produced by varying the probability threshold $t$.

\begin{figure*}[!tbp]
\centerline{\includegraphics[scale=0.8]{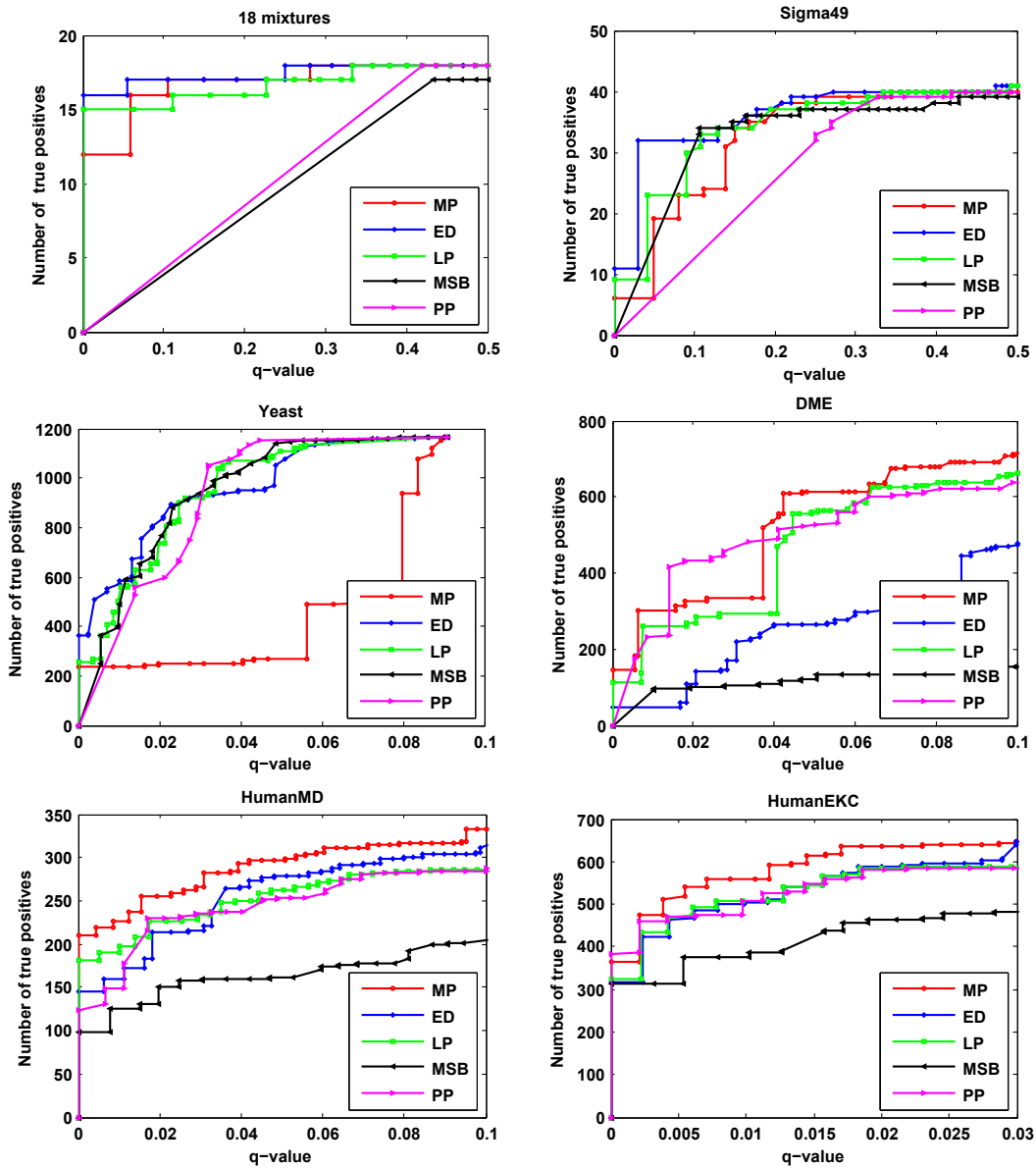}}
\caption{Identification performance comparison among MSBayesPro (MSB), ProteinProphet (PP) and our own three methods: multiple counting (MP), equal division (ED) and linear programming (LP). We only plot the curve up to 0.1 along the $x$-axis for yeast, DME and HumanMD datasets since people are particularly interested in the performance of different algorithms when the $q$-value or FDR is very small. For HumanEKC, the maximum $q$-value is $<0.04$ so that we choose 0.03 as the limit of $x$-axis. We cannot set the $q$-value range very small for 18 mixtures and Sigma49 datasets since the probabilities of top-scoring proteins in the several algorithms are all equal to one, hence we have to ship these proteins with same probabilities and then calculate the $q$-value of the first appearing protein with a different probability. }
\label{fig:qvalue}
\end{figure*}
Figure~\ref{fig:qvalue} plots the number of TPs identified by five methods at different $q$-values. It shows that our methods are competitive with available protein inference algorithms.  Throughout six datasets, our three methods can always achieve zero FPs among the highest ranking proteins while other two algorithms don't have such a property. This fact indicates that our methods have a strong distinction power of protein scores. More specifically, we have the following important observations.

First, the multiple counting method performs the best on DME, HumanMD and HumanEKC datasets. For DME and HumanMD, it reports the largest number of TPs under zero $q$-value. For HumanEKC, it just identifies 17 less proteins than ProteinProphet when $q$-value=0. Even though the multiple counting method doesn't keep such excellent performance on 18 mixtures and Sigma49 datasets, it doesn't perform the worst.

Second, equal division is the best performer (or tied with other algorithms) on 18 mixtures, Sigma49 and yeast datasets.  Similarly, under zero $q$-value, it identifies the most TPs on 18 mixtures, Sigma49 and yeast datasets. For DME, HumanMD and HumanEKC datasets, equal division doesn't have the worst performance as well. It beats at least one algorithm on DME, HumanMD and HumanEKC datasets.

Third, the LP model exhibits the most stable identification performance among these five methods. More precisely, its performance is at least the third best across all the five datasets. Other four algorithms cannot achieve such a property and they perform worse than at least three algorithms on some datasets. The number of these datasets is 2, 1, 2, 4 for multiple counting, equal division, ProteinProphet and MSBayesPro, respectively.
\begin{figure*}[!tbp]
\centerline{\includegraphics[scale=0.8]{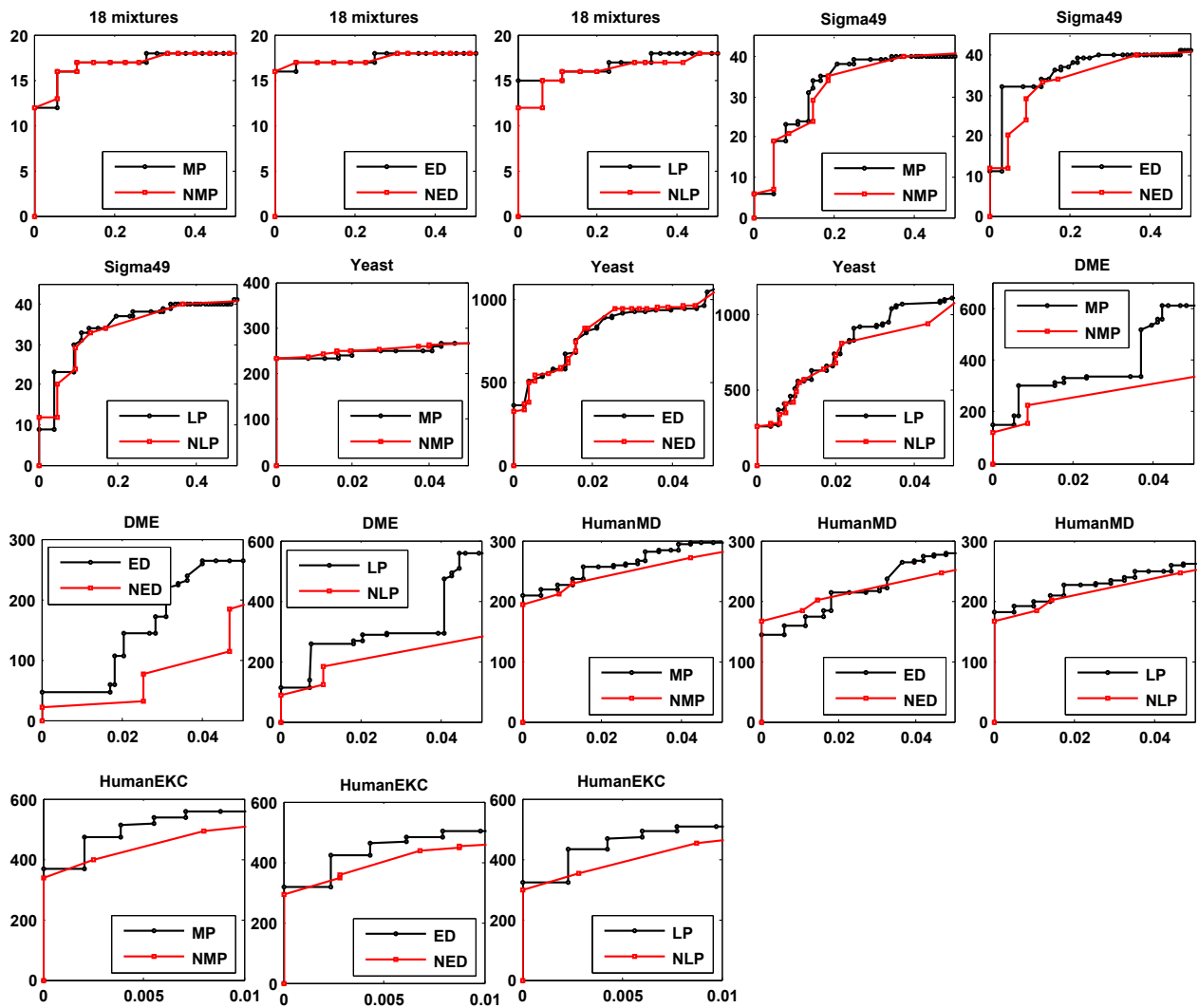}}
\caption{Identification performance comparison between the generalized spectral counting methods (MP, ED, LP) and the traditional spectral counting methods (NMP, NED, NLP). The $y$-axis is the number of true positives and  $x$-axis is the corresponding $q$-value (the minimum FDR to report these proteins). The abbreviations for different methods are the same as those in Figure~\ref{fig:qvalue}.}
\label{fig:ScoreProbability}
\end{figure*}

In the calculation of protein abundance, we generalize the number of MS/MS spectra to the sum of PSM probabilities. We wish such an extension may help us to distinguish proteins with the same number of PSMs and further improve the identification performance. To show this fact, Figure~\ref{fig:ScoreProbability} describes the performance gain when the generalized spectral counting is used instead of the traditional spectral counting. The experimental results of these three methods on the six datasets agree with our expectation: using the sum of PSM probabilities actually performs better than using the number of PSMs.

\begin{figure*}[!tbp]
\centerline{\includegraphics[scale=0.8]{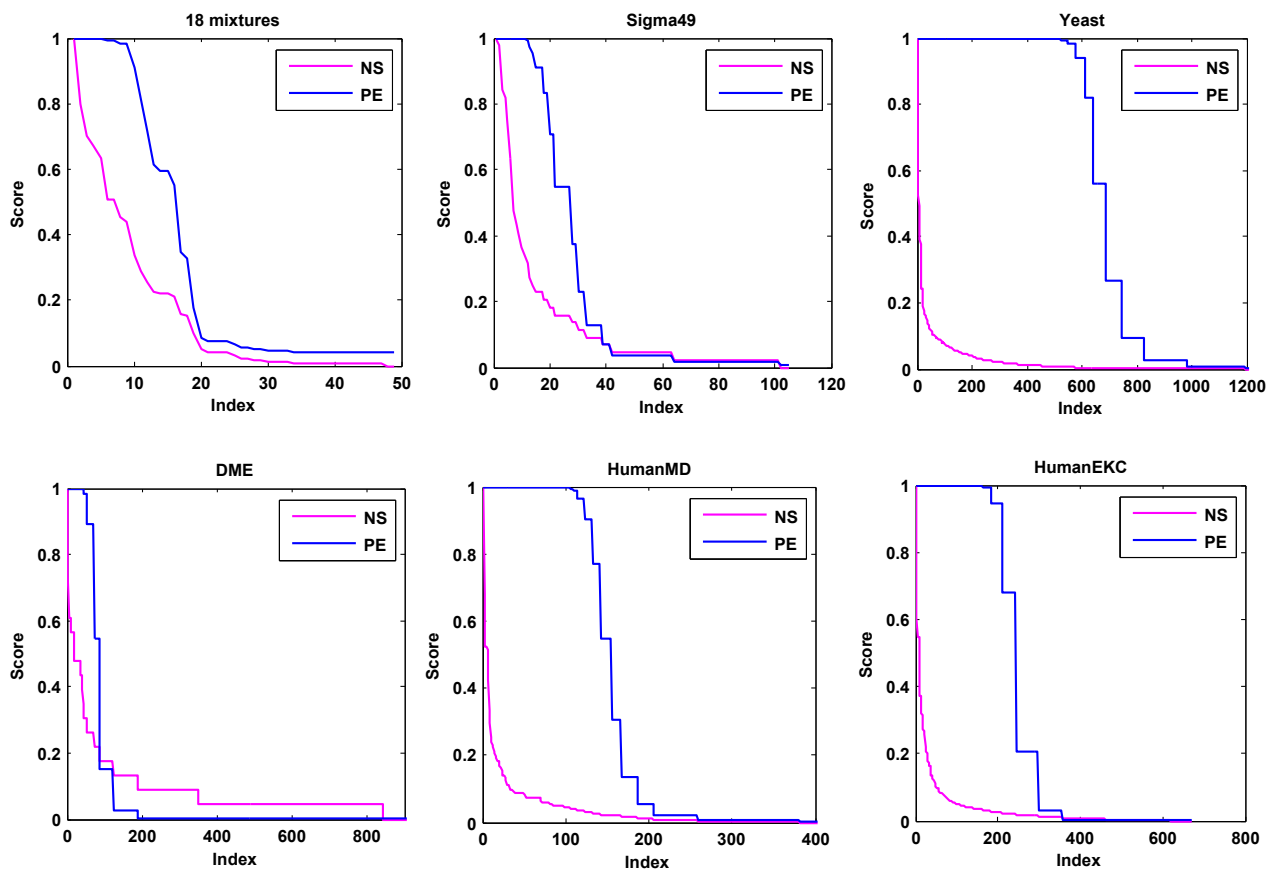}}
\caption{Comparison of the score distribution between normalized score (NS) and probability estimation (PE) when the protein abundance value is generated with the LP model. The scores of all the identified proteins are sorted by descending order.}
\label{fig:ScoreDistribution}
\end{figure*}

After obtaining the protein abundance, we use an EM algorithm to convert the abundance score into a well-calibrated probability. Alternatively, we can just normalize this abundance by dividing the maximum of all calculated protein abundances. This way gives us a protein score between 0 and 1 as well and keeps the holistic distribution of original protein abundance unchanged. Figure~\ref{fig:ScoreDistribution} shows the reason why we adopt the more complex probability estimation approach. It compares the distribution of normalized score and estimated probability using protein abundance calculated by LP model. For each of the six datasets, the area under the probability estimation curve is larger than that under the normalized score curve. It indicates that the probability estimation has a more uniform distribution than normalized protein score. Furthermore, the estimated probabilities of top-ranking proteins are very close to one but not equal to 1 so as to allow for distinction on a fine level.

\section{Conclusion}

Protein inference and protein quantification have been considered as two individual computational problems for a long time. In this paper, we investigate the feasibility of solving protein inference problem with existing protein quantification methods in the context of label-free protein quantification. The experimental results show that such a new angle enables us to obtain better identification performance even with some very simple quantification approaches available in the literature.

We have tested three spectral counting methods for solving the protein inference problem. These three methods can achieve good performance but none of them are consistently the best method on all the datasets. Thus, it is still necessary to develop better algorithms. In the future work, we plan to try more quantification methods to check if we can further improve the identification performance.

\section*{Acknowledgements}
This work was partially supported by the Natural Science Foundation of China under Grant No. 61003176.

\bibliographystyle{IEEEtran}
\bibliography{ProteinLT}

\begin{thebibliography}{10}
\providecommand{\url}[1]{#1}
\csname url@samestyle\endcsname
\providecommand{\newblock}{\relax}
\providecommand{\bibinfo}[2]{#2}
\providecommand{\BIBentrySTDinterwordspacing}{\spaceskip=0pt\relax}
\providecommand{\BIBentryALTinterwordstretchfactor}{4}
\providecommand{\BIBentryALTinterwordspacing}{\spaceskip=\fontdimen2\font plus
\BIBentryALTinterwordstretchfactor\fontdimen3\font minus
  \fontdimen4\font\relax}
\providecommand{\BIBforeignlanguage}[2]{{%
\expandafter\ifx\csname l@#1\endcsname\relax
\typeout{** WARNING: IEEEtran.bst: No hyphenation pattern has been}%
\typeout{** loaded for the language `#1'. Using the pattern for}%
\typeout{** the default language instead.}%
\else
\language=\csname l@#1\endcsname
\fi
#2}}
\providecommand{\BIBdecl}{\relax}
\BIBdecl

\bibitem{Analysis}
A.~I. Nesvizhskii, O.~Vitek, and R.~Aebersold, ``Analysis and validation of
  proteomic data generated by tandem mass spectrometry,'' \emph{Nature
  Methods}, vol.~4, no.~10, pp. 787--797, 2007.

\bibitem{Review}
T.~Huang, J.~Wang, W.~Yu, and Z.~He, ``Protein inference: A review,''
  \emph{Briefings in Bioinformatics}, vol.~13, no.~5, pp. 586--614, 2012.

\bibitem{Nikolov}
M.~Nikolov, C.~Schmidt, and H.~Urlaub, ``Quantitative mass spectrometry-based
  proteomics: an overview,'' \emph{Methods in Molecular Biology}, vol. 893, pp.
  85--100, 2012.

\bibitem{DTASelect}
D.~L. Tabb, H.~McDonald, and J.~R. Yates, ``{DTAS}elect and {C}ontrast: {t}ools
  for assembling and comparing protein identifications from shotgun
  proteomics,'' \emph{Journal of Proteome Research}, vol.~1, no.~1, pp. 21--26,
  2002.

\bibitem{Dbparser}
X.~Yang, V.~Dondeti, R.~Dezube, D.~M. Maynard, L.~Y. Geer, J.~Epstein, X.~Chen,
  S.~P. Markey, and J.~A. Kowalak, ``D{BP}arser: Web-based software for shotgun
  proteomic data analyses,'' \emph{Journal of Proteome Research}, vol.~3,
  no.~5, pp. 1002--1008, 2004.

\bibitem{Proteomicbing}
B.~Zhang, M.~C. Chambers, and D.~L. Tabb, ``Proteomic parsimony through
  bipartite graph analysis improves accuracy and transparency,'' \emph{Journal
  of Proteome Research}, vol.~6, no.~9, pp. 3549--3557, 2007.

\bibitem{Improvedze}
Z.-Q. Ma, S.~Dasari, M.~C. Chambers, M.~Litton, S.~M. Sobecki, L.~Zimmerman,
  P.~J. Halvey, B.~Schilling, P.~M. Drake, B.~W. Gibson, and D.~L. Tabb,
  ``I{DP}icker 2.0: {I}mproved protein assembly with high discrimination
  peptide identification filtering,'' \emph{Journal of Proteome Research},
  vol.~8, no.~8, pp. 3872--3881, 2009.

\bibitem{Markey}
D.~J. Slotta, M.~A. McFarland, and S.~P. Markey, ``Mass{S}ieve: Panning {MS/MS}
  peptide data for proteins,'' \emph{Proteomics}, vol.~10, no.~16, pp.
  3035--3039, 2010.

\bibitem{Aseon}
A.~I. Nesvizhskii, A.~Keller, E.~Kolker, and R.~Aebersold, ``A statistical
  model for identifying proteins by tandem mass spectrometry,''
  \emph{Analytical Chemistry}, vol.~75, no.~17, pp. 4646--4658, 2003.

\bibitem{Ebpthomas}
T.~S. Price, M.~B. Lucitt, W.~Wu, D.~J. Austin, A.~Pizarro, A.~K. Yocum, I.~A.
  Blair, G.~A. FitzGerald, and T.~Grosser, ``{EBP}: Protein identification
  using multiple tandem mass spectrometry datasets,'' \emph{Molecular \&
  Cellular Proteomics}, vol.~6, no.~3, pp. 527--536, 2007.

\bibitem{Probabilityfeng}
J.~Feng, D.~Q. Naiman, and B.~Cooper, ``Probability model for assessing
  proteins assembled from peptides sequences inferred from tandem mass
  spectrometry data,'' \emph{Analytical Chemistry}, vol.~79, no.~10, pp.
  3901--3911, 2007.

\bibitem{Aweaherly}
D.~B. Weatherly, J.~A. Atwood, T.~A. Minning, C.~Cavola, R.~L. Tarleton, and
  R.~Orlando, ``A heuristic method for assigning a false-discovery rate for
  protein identifications from mascot database search results,''
  \emph{Molecular \& Cellular Proteomics}, vol.~4, no.~6, pp. 762--772, 2005.

\bibitem{Ayong}
Y.~F. Li, R.~J. Arnold, Y.~Li, P.~Radivojac, Q.~Sheng, and H.~Tang, ``A
  {B}ayesian approach to protein inference problem in shotgun proteomics,''
  \emph{Journal of Computational Biology}, vol.~16, no.~8, pp. 1--11, 2009.

\bibitem{Efficientoliver}
O.~Serang, M.~J. MacCoss, and W.~S. Noble, ``Efficient marginalization to
  compute protein posterior probabilities from shotgun mass spectrometry
  data,'' \emph{Journal of Proteome Research}, vol.~9, no.~10, pp. 5346--5357,
  2010.

\bibitem{Anmoore}
R.~E. Moore, M.~K. Young, and T.~D. Lee, ``Qscore: an algorithm for evaluating
  sequest database search results,'' \emph{Journal of American Society for Mass
  Spectrometry}, vol.~13, no.~4, pp. 378--386, 2002.

\bibitem{Statisticalr}
R.~G. Sadygov, H.~Liu, and J.~R. Yates, ``Statistical models for protein
  validation using tandem mass spectral data and protein amino acid sequence
  databases,'' \emph{Analytical Chemistry}, vol.~76, no.~6, pp. 1664--1671,
  2004.

\bibitem{Improved}
M.~Bern and D.~Goldberg, ``Improved ranking functions for protein and
  modification-site identifications,'' \emph{Journal of Computational Biology},
  vol.~15, no.~7, pp. 705--719, 2008.

\bibitem{Aqunhua}
Q.~Li, M.~MacCoss, and M.~Stephens, ``A nested mixture model for protein
  identification using mass spectrometry,'' \emph{The Annals of Applied
  Statistics}, vol.~4, no.~2, pp. 962--987, 2010.

\bibitem{Achangyu}
C.~Shen, Z.~Wang, G.~Shankar, X.~Zhang, and L.~Li, ``A hierarchical statistical
  model to assess the confidence of peptides and proteins inferred from tandem
  mass spectrometry,'' \emph{Bioinformatics}, vol.~24, no.~2, pp. 202--208,
  2008.

\bibitem{Direct}
M.~Spivak, J.~Weston, M.~J. MacCoss, and W.~S. Noble, ``Direct maximization of
  protein identifications from tandem mass spectra,'' \emph{Molecular \&
  Cellular Proteomics}, vol.~11, no.~2, p. M111.012161, 2012.

\bibitem{Searle}
B.~C. Searle, ``Scaffold: A bioinformatic tool for validating {MS/MS}-based
  proteomic studies,'' \emph{Proteomics}, vol.~10, no.~6, pp. 1265--1269, 2010.

\bibitem{Azeng}
Z.~He, C.~Yang, and W.~Yu, ``A partial set covering model for protein mixture
  identification using mass spectrometry data,'' \emph{IEEE/ACM Transactions on
  Computational Biology and Bioinformatics}, vol.~8, no.~2, pp. 368--380, 2011.

\bibitem{Improvingbingwen}
B.~Lu, A.~Motoyama, C.~Ruse, J.~Venable, and J.~R. Yates, ``Improving protein
  identification sensitivity by combining {MS} and {MS/MS} information for
  shotgun proteomics using {LTQ-O}rbitrap high mass accuracy data,''
  \emph{Analytical Chemistry}, vol.~80, no.~6, pp. 2018--2025, 2008.

\bibitem{Proteinpaul}
P.~Kearney, H.~Butler, K.~Eng, and P.~Hugo, ``Protein identification and
  peptide expression resolver: Harmonizing protein identification with protein
  expression data,'' \emph{Journal of Proteome Research}, vol.~7, no.~1, pp.
  234--244, 2008.

\bibitem{Ramakrishnan1}
S.~R. Ramakrishnan, C.~Vogel, T.~Kwon, L.~O. Penalva, E.~M. Marcotte, and D.~P.
  Miranker, ``Mining gene functional networks to improve mass-spectrometry
  based protein identification,'' \emph{Bioinformatics}, vol.~25, no.~22, pp.
  2955--2961, 2009.

\bibitem{Networkjing}
J.~Li, L.~J. Zimmerman, B.-H. Park, D.~L. Tabb, D.~C. Liebler, and B.~Zhang,
  ``Network-assisted protein identification and data interpretation in shotgun
  proteomics,'' \emph{Molecular Systems Biology}, vol.~5, p. 303, 2009.

\bibitem{Ramakrishnan2}
S.~R. Ramakrishnan, C.~Vogel, J.~T. Prince, R.~Wang, Z.~Li, L.~O. Penalva,
  M.~Myers, E.~M. Marcotte, and D.~P. Miranker, ``Integrating shotgun
  proteomics and m{RNA} expression data to improve protein identification,''
  \emph{Bioinformatics}, vol.~25, no.~11, pp. 1397--1403, 2009.

\bibitem{Grobei}
M.~A. Grobei, E.~Qeli, E.~Brunner, H.~Rehrauer, R.~Zhang, B.~Roschitzki,
  K.~Basler, C.~H. Ahrens, and U.~Grossniklaus, ``Deterministic protein
  inference for shotgun proteomics data provides new insights into
  {A}rabidopsis pollen development and function,'' \emph{Genome Research},
  vol.~19, no.~10, pp. 1786--1800, 2009.

\bibitem{Douglas}
E.~Qeli and C.~H. Ahrens, ``Peptide{C}lassifier for protein inference and
  targeted quantitative proteomics,'' \emph{Nature Biotechnology}, vol.~28,
  no.~7, pp. 647--650, 2010.

\bibitem{Proteinsarah}
S.~Gerster, E.~Qeli, C.~H. Ahrens, and P.~Buhlmann, ``Protein and gene model
  inference based on statistical modeling in k-partite graphs,''
  \emph{Proceedings of the National Academy of Sciences of USA}, vol. 107,
  no.~27, pp. 12\,101--12\,106, 2010.

\bibitem{William}
W.~M. Old, K.~Meyer-Arendt, L.~Aveline-Wolf, K.~G. Pierce, A.~Mendoza, J.~R.
  Sevinsky, K.~A. Resing, and N.~G. Ahn, ``Comparison of label-free methods for
  quantifying human proteins by shotgun proteomics,'' \emph{Molecular \&
  Cellular Proteomics}, vol.~4, pp. 1487--1502, 2005.

\bibitem{Florens}
Y.~Zhang, Z.~Wen, M.~P. Washburn, and L.~Florens, ``Refinements to label free
  proteome quantitation: How to deal with peptides shared by multiple
  proteins,'' \emph{Molecular \& Cellular Proteomics}, vol.~82, no.~6, pp.
  2272--2281, 2010.

\bibitem{Recomb2009SharedPeptide}
B.~Dost, N.~Bandeira, X.~Li, Z.~Shen, S.~Briggs, and V.~Bafna, ``Shared
  peptides in mass spectrometry based protein quantification,'' in
  \emph{Research in Computational Molecular Biology}, vol. 5541, 2009, pp.
  356--371.

\bibitem{Muralidharan}
K.~A. Neilson, N.~A. Ali, S.~Muralidharan, M.~Mirzaei, M.~Mariani,
  G.~Assadourian, A.~Lee, S.~C. van Sluyter, and P.~A. Haynes, ``Less label,
  more free: {A}pproaches in label-free quantitative mass spectrometry,''
  \emph{Proteomics}, vol.~11, no.~4, pp. 535--553, 2011.

\bibitem{Linfeng}
D.~H. Lundgren, S.-I. Hwang, L.~Wu, and D.~K. Han, ``Role of spectral counting
  in quantitative proteomics,'' \emph{Expert {R}eview of {P}roteomics}, vol.~7,
  no.~1, pp. 39--53, 2010.

\bibitem{Fermin}
H.~Choi, D.~Fermin, and A.~I. Nesvizhskii, ``Significance analysis of spectral
  count data in label-free shotgun proteomics,'' \emph{Molecular \& Cellular
  Proteomics}, vol.~7, no.~12, pp. 2373--2385, 2008.

\bibitem{Probability}
D.~N. Perkins, D.~J.Pappin, D.~M.Creasy, and J.~S. Cottrell,
  ``Probability-based protein identification by searching sequence databases
  using mass spectrometry data,'' \emph{Electrophoresis}, vol.~20, no.~18, pp.
  3551--3567, 1999.

\bibitem{Aebersold3}
A.~Keller, A.~I. Nesvizhskii, E.~Kolker, and R.~Aebersold, ``Empirical
  statistical model to estimate the accuracy of peptide identifications made by
  {MS/MS} and database search,'' \emph{Analytical Chemistry}, vol.~74, no.~20,
  pp. 5383--5392, 2002.

\bibitem{Usaite}
R.~Usaite, J.~Wohlschlegel, J.~D. Venable, S.~K. Park, J.~Nielsen, L.~Olsson,
  and J.~R. Yates, ``Characterization of global yeast quantitative proteome
  data generated from the wild-type and glucose repression saccharomyces
  cerevisiae strains: The comparison of two quantitative methods,''
  \emph{Journal of Proteome Research}, vol.~7, no.~1, pp. 266--275, 2008.

\bibitem{Zybailov}
B.~Zybailov, H.~Rutschow, G.~Friso, A.~Rudella, O.~Emanuelsson, Q.~Sun, and
  K.~J. van Wijk, ``Sorting signals, {N}-terminal modifications and abundance
  of the chloroplast proteome,'' \emph{PLoS ONE}, vol.~3, no.~4, p. e1994,
  2008.

\bibitem{JingGao}
J.~Gao and P.-N. Tan, ``Converting output scores from outlier detection
  algorithms into probability estimates,'' in \emph{IEEE International
  Conference on Data Mining}, Hong Kong, China, December 2006, pp. 212--221.

\bibitem{Platt}
J.~C. Platt, ``Probabilistic outputs for support vector machines and comparison
  to regularized likelihood methods,'' in \emph{Advances in Large Margin
  Classifiers}.\hskip 1em plus 0.5em minus 0.4em\relax MIT Press, 2000, pp.
  61--74.

\bibitem{18mixtures}
J.~Klimek, J.~S. Eddes, and L.~Hohmann, ``The {S}tandard {P}rotein {M}ix
  {D}atabase: {A} diverse data set to assist in the production of improved
  peptide and protein identification software tools,'' \emph{Journal of
  Proteome Research}, vol.~7, no.~1, pp. 96--103, 2008.

\bibitem{DMEBrunner}
E.~Brunner, C.~H. Ahrens, S.~Mohanty, H.~Baetschmann, S.~Loevenich,
  F.~Potthast, E.~W. Deutsch, C.~Panse, U.~de~Lichtenberg, O.~Rinner, H.~Lee,
  P.~G.~A. Pedrioli, J.~Malmstrom, K.~Koehler, S.~Schrimpf, J.~Krijgsveld,
  F.~Kregenow, A.~J.~R. Heck, E.~Hafen, R.~Schlapbach, and R.~Aebersold, ``A
  high-quality catalog of the drosophila melanogaster proteome,'' \emph{Nature
  Biotechnology}, vol.~25, no.~5, pp. 576--583, 2007.

\bibitem{Unimod}
C.~M. David and J.~S. Cottrell, ``Unimod: Protein modifications for mass
  spectrometry,'' \emph{Proteomics}, vol.~4, no.~6, pp. 1534--1536, 2004.

\end{thebibliography}

\end{document}